\documentclass[prb,twocolumn,showpacs,superscriptaddress]{revtex4-1}
\usepackage{epsfig}
\usepackage{graphicx}
\usepackage{amsfonts}
\usepackage[figuresright]{rotating}
\usepackage{amssymb}
\usepackage{amsmath}
\usepackage{psfrag}
\usepackage[colorlinks,linkcolor=blue,anchorcolor=blue,citecolor=blue,urlcolor=blue]{hyperref}
\newtheorem{theorem}{Theorem}
\newtheorem{lemma}{Lemma}
\newtheorem{corollary}{Corollary}
\newenvironment{remark}[1][Remark]{\noindent \textbf{#1: }}{\ \rule{0.5em}{0.5em}}

\newenvironment{proof}[1][Proof]{\noindent \textbf{#1: }}{\ \rule{0.5em}{0.5em}}

\def\avg#1{\langle#1\rangle}

\def\be{\begin{equation}} \def\ee{\end{equation}}
\def\bea{\begin{eqnarray}} \def\eea{\end{eqnarray}}

\def\nn{\nonumber}
\def\pp{\parallel}

\newcommand{\ket}[1]{| #1 \rangle}

\begin{document}
\title{Exact results for itinerant ferromagnetism
in a $t_{2g}$ orbital system on cubic and square lattices}
\author{Yi Li}
\affiliation{Princeton Center for Theoretical Science, Princeton
 University, Princeton, NJ 08544}
\date {March 12, 2015}

\begin{abstract}
We study itinerant ferromagnetism in a $t_{2g}$ multiorbital Hubbard
system in the cubic lattice, which consists of three planar oriented
orbital bands of $d_{xy}$, $d_{yz}$, and $d_{zx}$.
Electrons in each orbital band can only move within a two-dimensional plane
in the three-dimensional lattice parallel to the corresponding orbital orientation.
Electrons of different orbitals interact through
the on-site multiorbital interactions including Hund's coupling.
The strong coupling limit is considered in which there are no doubly
occupied orbitals but multiple on-site occupations are allowed.
We show that in the case in which there is one and only one hole
for each orbital band in each layer parallel to the orbital orientation,
the ground state is a fully spin-polarized itinerant ferromagnetic state,
which is unique apart from the trivial spin degeneracy.
When the lattice is reduced into a single two-dimensional layer, the $d_{zx}$ and
$d_{yz}$ bands become quasi-one-dimensional while the $d_{xy}$ band remains two-dimensional.
The ground state ferromagnetism also appears in the strong coupling limit
as a generalization of the double exchange mechanism.
Possible applications to the systems of SrRuO$_3$ and LaAlO$_3$/SrTiO$_3$ interface are discussed.
\end{abstract}
\maketitle

\section{Introduction}
\label{sec:intro}

Itinerant ferromagnetism (FM) is not only a representative
strong-correlation problem, but also a highly non-perturbative one
\cite{lieb1962,mattis2006,nagaoka1966,roth1966,kugel1973,
hertz1976,moriya1985,torrance1987,gill1987,hirsch1989,shastry1990,
mielke1991,mielke1991a,tasaki1992, millis1993,belitz2005,lohneysen2007,liu2012}.
It is widely known as a long-standing problem of condensed matter physics,
and also a current research focus in ultra-cold
atom physics \cite{duine2005,Jo2009,zhangSZ2010,berdnikov2009,
pekker2011a,chang2010,cui2014,pilati2014}.
The Stoner mechanism states that polarized electron systems can save
the exchange interaction energy.
Nevertheless, because of the associated cost of kinetic energy,
FM is not guaranteed even in the presence of very strong
repulsions.
For example, in rigorously one-dimensional (1D) systems, no matter
how strong the repulsive interactions are, the ground state is always a spin singlet, which is
known as the famous Lieb-Mattis theorem \cite{lieb1962}.
In other words, electrons can remain unpolarized but avoid
each other to reduce interaction, nevertheless, their wave functions
are strongly-correlated.
Certainly the Lieb-Mattis theorem in 1D only applies for spin-independent systems. Ferromagnetism in 1D is still possible if the interaction is spin-dependent.

Because of the strong correlation nature of itinerant FM, exact theorems
are important to provide reference points.
Nagaoka's theorem is an early example, which applies to
the infinite $U$ Hubbard model with a single hole
in the half-filled background \cite{nagaoka1966,tasaki1989,tian1991}.
The fully polarized FM state facilitates the hole's coherent
motion, which minimizes the kinetic energy of the hole
and is therefore selected as the ground state.
Another class of FM theorems is based on the flat band structure
on line graphs \cite{mielke1991,mielke1991a,mielke1992,mielke1993,
tasaki1992}.
Because of the divergence of density of states in the flat band, the
kinetic energy cost because of spin polarization is suppressed.
Metallic FM states with a dispersive band structure
have also been proved \cite{Tasaki2003,Tanaka2007}.

Recently, a ground state FM theorem has been proved in both two-dimensional (2D) square
and three-dimensional (3D) cubic lattices systems with multiorbital structures \cite{li2014}.
The band structure behaves like decoupled orthogonal 1D chains;
while, different chains are coupled at their crossing site
through multiorbital Hubbard interactions.
In particular,  spins of each chain are not conserved but
coupled by Hund's interaction. Hence, the ground state FM ordering is
genuinely 2D or 3D.
Different from Nagaoka's theorem, the result of multiorbital FM allows a stable FM
phase over a large region of filling factors in both 2D and 3D.
An important consequence of this theorem is that
the sign structure of the many-body Hamiltonian matrix
leads to the absence of the quantum Monte-Carlo (QMC) sign problem
\cite{li2014}.
Consequently, QMC simulations on finite temperature thermodynamic properties
of itinerant FM have been performed \cite{xuSL2014}, which yield
asymptotically exact results and shed light on the mechanism of
magnetic phase transitions in the strong-coupling limit.

In this article, we generalize Nagaoka's theorem of itinerant FM
from the single orbital system to multiorbital systems.
We consider the 3D cubic lattice and each site consists of three
$t_{2g}$ orbitals: $d_{xy}$, $d_{yz}$, and $d_{zx}$.
Each orbital has a planar orientation, and the associated band structure
is quasi-2D like.
Electrons of different orbitals interact through the on-site multiorbital
interactions including Hund's coupling.
In the limit of intra orbital interaction $U \rightarrow \infty$,
states with doubly occupied orbitals are projected out.
When each plane of the cubic lattice has one and only one hole
in the in-plane orbital band, this system can be be viewed as
crossing layers of Nagaoka FM states.
We prove that, in this limit, the ground state of this system is
the fully spin-polarized itinerant ferromagnetic state, and
it is non-degenerate apart from the trivial spin degeneracy.
Furthermore, when this system is reduced to a single 2D layer system
of $t_{2g}$ orbitals, the $d_{zx}$ and $d_{yz}$ orbital bands become
quasi-1D and coupled to the quasi-2D band of $d_{xy}$ through Hund's coupling.
The ground state FM is still valid,
where the quasi-1D $d_{zx}$ and $d_{yz}$ bands are allowed to take general
values of filling, while, the $d_{xy}$ band can possess a single hole or
be fully filled.
Although the above exact results require an idealized strong-coupling limit,
the strong correlation physics that they
imply sheds important light on the mechanism of
itinerant FM in transition metal oxides.


The rest of this paper is organized as follows:
In Sec. \ref{sect:model}, the multiorbital Hubbard model
for the $t_{2g}$ orbital in the 3D cubic lattice is defined.
In Sec. \ref{sect:3d}, Theorem 1 of the ground state itinerant FM
in the 3D $t_{2g}$ orbital system is proved.
In Sec. \ref{sect:layer}, Theorem 2 of the ground state
itinerant FM for the reduced 2D layered system is proved.
Discussion on connections to experiment systems is
provided in Sec. \ref{sect:exp}.
Conclusions are presented in Sec. \ref{sect:cons}.

\section{The model Hamiltonian: a 3D multiorbital Hubbard model}
\label{sect:model}

In this section, we define a 3D multiorbital Hubbard model in the
3D cubic lattice, which will be shown to possess itinerant
FM ground states under conditions (I) and (II) in Sec. \ref{sect:3d}.

\begin{figure}[tbp]
\centering\epsfig{file=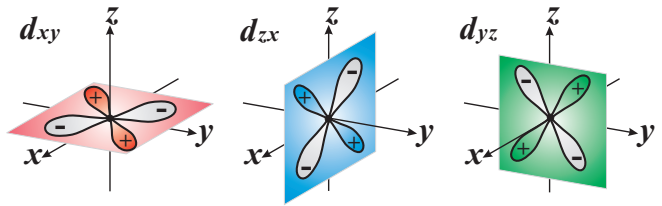,clip=1,width=\linewidth,angle=0}
\caption{The Wannier orbital wavefunctions of $t_{2g}$ orbitals:
$d_{xy}$, $d_{yz}$ and $d_{zx}$.
For electrons in the $d_{a}$ orbitals $(a=xy,yz,zx)$, they can only move
along the $xy$-, $yz$-, or $zx$-plane, respectively,
but not perpendicular to the orbital orientation plane.
}
\label{fig:orbital}
\end{figure}

We consider a $t_{2g}$ orbital system filled with spin -$1/2$ electrons;
{\it i.e.}, each site possesses $d_{xy}$, $d_{yz}$ and $d_{zx}$ orbitals.
The Wannier wavefunction of the $t_{2g}$ orbitals is planar-like
as shown in Fig. \ref{fig:orbital}.
The kinetic energy of each orbital band exhibits a 2D structure:
Say, for electrons in the $d_{xy}$ orbital, they can only move in the
$xy$-plane with a hopping amplitude $t_\pp$.
However, their hopping amplitude $t_\perp$ along the transverse direction
of the $z$ axis is very small.
Usually, the in-plane hopping $t_\pp$ is assisted by the $p$ orbitals of
oxygen anions lying at the middle point of the bond, which leads to large hopping amplitudes;
while, the transverse hopping $t_\perp$ can only be attributed to the direct
overlap between two $d_{xy}$ orbitals offset along the $z$ axis.
Since $d$ orbital Wannier functions are nearly localized and the $z$ axis
is perpendicular to the orbital plane, $t_\pp$ is negligible in
realistic transition metal oxides.
Similarly, electrons in the $d_{yz}$ and $d_{zx}$ orbitals only hop
along the $yz$ and $zx$-planes, respectively.

Because of the different parity eigenvalues of these three $t_{2g}$ orbitals
with respect to the $xy$, $yz$, and $zx$-planes, they do not hybridize
by the nearest neighbor hopping.
If we neglect the longer range hopping terms, the kinetic energy part can
simply be written as
\bea
H^K=H^K_{xy} +H^K_{yz} +H^K_{zx},
\eea
where $H^K_{xy}$,  $H^K_{yz}$, and $H^K_{zx}$ are the kinetic energies
of electrons in the $xy$, $yz$, and $zx$ orbital bands, respectively.
The kinetic energy for the $xy$ orbital band  is expressed as
\bea
H^K_{xy}&=&\sum_{\mathbf{r}} t_\pp \Big(d^\dagger_{xy,\sigma}(\mathbf{r})
d_{xy,\sigma}(\mathbf{r}+a_0\hat x)\nn \\
&+&d^\dagger_{xy,\sigma}(\mathbf{r}) d_{xy,\sigma}(\mathbf{r}+a_0 \hat y)+h.c.
\Big),
\label{eq:kin}
\eea
where $a_0$ is the lattice constant;
$d_{xy,\sigma}(\mathbf{r})$ is the annihilation operator in the
$d_{xy}$ orbital on site $\mathbf{r}$ with the spin-index
$\sigma=\uparrow$ or $\downarrow$.
For convenience later, we choose $t_\pp$ positive.
For the bipartite lattice such as the cubic one, the sign of
$t_\pp$ can be flipped by a gauge transformation, which does
not affect any physical observable.
The transverse hopping $t_\perp$ term is neglected in Eq. (\ref{eq:kin}).
Similarly, $H^K_{yz (zx)}$ can also be defined by permuting
the indices of orbitals and hopping directions in  $H^K_{xy}$,
whose expressions are not repeated here.

The interaction term is the standard multiorbital Hubbard interaction
\cite{roth1966,kugel1973,cyrot1975,oles1983} defined on-site as
\bea
H^I&=&U\sum_{\mathbf{r}, a} n_{a,\uparrow}(\mathbf{r}) n_{a,\downarrow} (\mathbf{r})
\nn \\
&-& J\sum_{\mathbf{r}, a\neq b}\left(\vec S_{a} (\mathbf{r})\cdot \vec S_b (\mathbf{r})
-\frac{1}{4} n_a (\mathbf{r}) n_b(\mathbf{r})\right)\nn \\
&+&V\sum_{\mathbf{r}, a\neq b} n_a (\mathbf{r}) n_b(\mathbf{r}) \nn \\
&+&\Delta\sum_{\mathbf{r}, a \neq b }
\left(d^\dagger_{a,\uparrow} (\mathbf{r})d^\dagger_{a,\downarrow} (\mathbf{r})
d_{b,\downarrow} (\mathbf{r}) d_{b,\uparrow} (\mathbf{r})
+h.c. \right),
\label{eq:Hint}
\eea
where $a=xy,yz,zx$ is the orbital index; $n_{a,\sigma}(\mathbf{r})$ is the
number of electrons occupying the $a$ orbital at site $\mathbf{r}$ with spin-index
$\sigma$; $n_a=n_{a,\uparrow}+n_{a,\downarrow}$;
$\vec S_a(\mathbf{r})$ is the spin operator of the $a$-th orbital at site $\mathbf{r}$.

Equation (\ref{eq:Hint}) contains all the possible terms satisfying
the spin SU(2) symmetry and the lattice cubic symmetry.
The $U$ term is the usual intra orbital Hubbard interaction;
the $V$ term is the inter orbital Hubbard interaction;
the $J$ term is Hund's coupling with $J>0$; and the
$\Delta$ term describes the singlet pairing hopping process among
different orbitals.
The expressions of $U$, $V$, $J$, and $\Delta$
are presented in Appendix \ref{appdx:parameter}
following the standard physical meaning of two-body Coulomb interactions.

\section{Ferromagnetism in the 3D $t_{2g}$ orbital system}
\label{sect:3d}

In this section, we consider the 3D $t_{2g}$ orbital systems
in the cubic lattice of size $L_x\times L_y\times L_z$.
We also assume the following two conditions, \\
(I) \textit{$U\rightarrow +\infty$,  $\Delta$ is finite}; \\
(II) \textit{For each orbital band, there is one and only one hole
in every layer parallel to the orbital plane. E.g., there is one and
only one hole in every $xy$-plane in the $d_{xy}$ orbital band,
and similarly for the $d_{yz}$ and $d_{zx}$ orbital bands}.\\

Condition II can be well-defined because of the following
lemma whose proof is obvious.
{\lemma
The Hamiltonian of Eqs.
(\ref{eq:kin}) and (\ref{eq:Hint}) conserves particle number
in each orbital band in each plane parallel to the orbital
orientation. }\\
Accordingly, the Hilbert space of the system can be
factorized as the tensor product of the Hilbert space of
each orbital band in each layer
as
\bea
{\cal H}=\bigotimes_{l_z=1}^{L_z}{\cal H}_{l_z}^{xy}
\bigotimes_{l_x=1}^{L_x}{\cal H}_{l_x}^{yz}
\bigotimes_{l_y=1}^{L_y}{\cal H}_{l_y}^{zx},
\eea
where $l_{z,x,y}$ are the indices of the $xy$, $yz$, and
$zx$-planes, respectively.
Under condition I, states with doubly occupied orbitals are
projected out, and each orbital can only be occupied at most by one particle.
Further, condition II restricts one and only one hole for each orbital band in a layer.
In each given Hilbert
space ${\cal H}^a_{l_i}$, each state is determined by the location of
the hole and the spin configuration at other sites.
For example, in the Hilbert space of the $d_{xy}$ orbital of the
$l_z$-th layer, we can label all the $d_{xy}$ orbitals of this layer
in an arbitrary order by the index $i_{l_z}=1,\cdots,L_x L_y$.
Then, the states in this layer can be represented as
\bea
|h^{xy}_{l_z};\{\sigma\}_{l_z}\rangle= (-)^{h^{xy}_{l_z}} {\prod_{i_{l_z}}}^\prime
d^\dagger_{xy,\sigma} (i_{l_z})|0\rangle,
\label{eq:sign}
\eea
where $h^{xy}_{l_z}$ labels the location of the hole;
$\{\sigma\}_{l_z}$ represents the spin configuration;
$\prod^\prime$ means the ordered product of the creation operators
except the one at the location of the hole,
$ \prod_{i_{l_z}}^\prime
d^\dagger_{xy,\sigma} (i_{l_z})|0\rangle
= d^\dagger_{xy,\sigma_1} (1) \cdots d^\dagger_{xy,\sigma_{h-1}} (h^{xy}_{l_z}-1) \widehat{d^\dagger_{xy,\sigma_{h}} (h^{xy}_{l_z})} d^\dagger_{xy,\sigma_{h+1}} (h^{xy}_{l_z}+1) \cdots d^\dagger_{xy,\sigma_{L_xL_y}} (L_xL_y)|0\rangle$ with the ``hat" means the operator below it does not appear.
Then, we can define the bases of the product Hilbert space for our entire system as
\bea
|\{h\}, \{\sigma\}\rangle=&&\bigotimes_{l_z=1}^{L_z} |h^{xy}_{l_z};\{\sigma\}_{l_z}\rangle\bigotimes_{l_x=1}^{L_x}  |h^{yz}_{l_x};\{\sigma\}_{l_x}\rangle\nn \\
&&\bigotimes_{l_y=1}^{L_y} |h^{zx}_{l_y};\{\sigma\}_{l_y}\rangle,
\label{eq:bases}
\eea
where, $\{h\}$ represents the locations of all the holes in a given state
and $\{\sigma\}$ represents the spin configuration of this state with the labels of orbitals
and layers omitted,.
Because of the spin conservation, we can decompose the Hilbert space
into different sectors ${\cal H}^{S_z}$ by the value of
the $z$ component of total spin $S_z$, denoted as ${\cal H}=\bigoplus {\cal H}^{S_z}$.
Nevertheless, ${\cal H}^{S_z}$ cannot be further factorized as the product space of
different orbital bands and layers.

Next, let us prove two lemmas as the preparation of the FM Theorem 1.
{\lemma {\bf (Non-positivity)}
Under the bases $|\{h\}, \{\sigma\}\rangle$ defined above
for the Hilbert space ${\cal H}$ with total spin $S_z$, the off-diagonal matrix
elements of the many-body Hamiltonian $H=H_{kin}+H_{int}$
(see Eqs. (\ref{eq:kin}) and (\ref{eq:Hint}))
are non-positive.}

\proof: The off-diagonal matrix elements are contributed by the hopping
part and Hund's interaction part.
The pairing hopping term does not exist in the limit of
$U\rightarrow +\infty$ since states with doubly occupied
orbitals have been projected out.
For the hopping term, because of the sign convention of the
many-body bases defined in Eq. (\ref{eq:bases}) inherited
from Eq. (\ref{eq:sign}), it is easy to check that
\bea
\avg{\{h\}, \{\sigma\}|H_t|\{h'\}, \{\sigma'\}}=-t \, \, \textrm{or} \, \, 0.
\eea
This step is the same as that in the proof of the usual Nagaoka theorem for
a 2D single orbital Hubbard model \cite{tasaki1989}:
Although here are ($L_x+L_y+L_z$) holes in our system,
the fermion ordering does not change under hopping because of Lemma 1.
For the $xy$ component of Hund's interaction
$H_{J_{xy}}=-J/2 \sum_{a \neq b}(S^+_a S^-_b + S^-_a S^+_b)$ with $S_a^{\pm}=S_a^x \pm i S_a^y$, it does not
change the fermion ordering either, and thus, its matrix elements read
\bea
\avg{\{h\}, \{\sigma\}|H_{J_{xy}}|\{h'\}, \{\sigma'\}}=-J/2 \, \, \textrm{or} \, \, 0,,
\eea
which are also non-positive.
The $V$ term and the $z$ component of Hund's interaction only
contribute to the diagonal part of the many-body matrix. \textit{Q.E.D.}

\begin{figure}[tbp]
\centering\epsfig{file=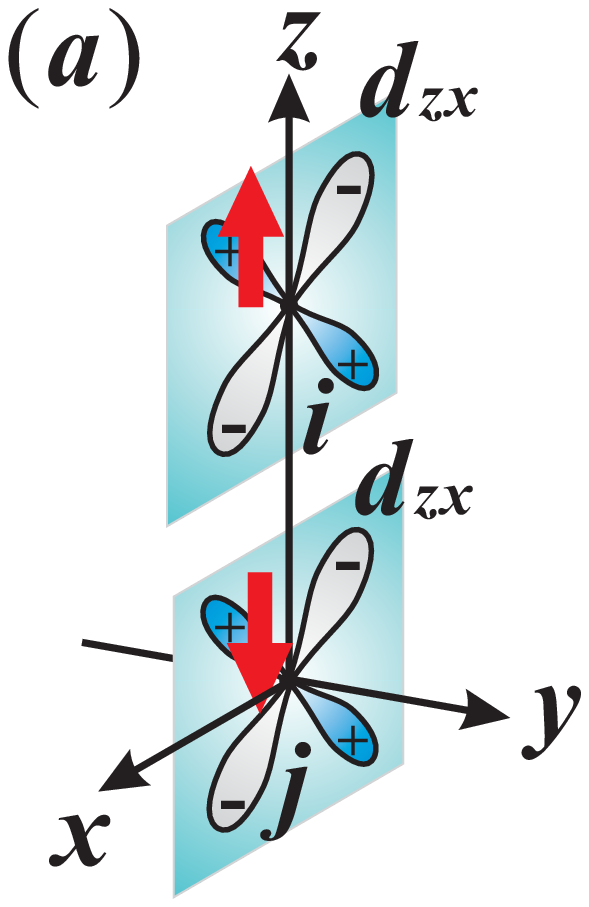,clip=1,width=0.35\linewidth,angle=0}
\centering\epsfig{file=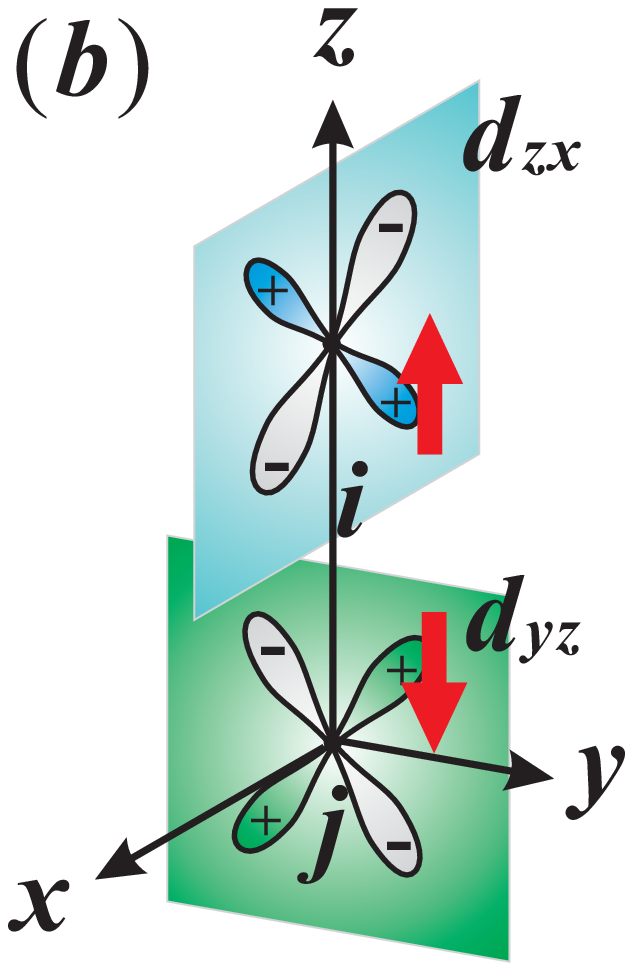,clip=1,width=0.35\linewidth,angle=0}
\centering\epsfig{file=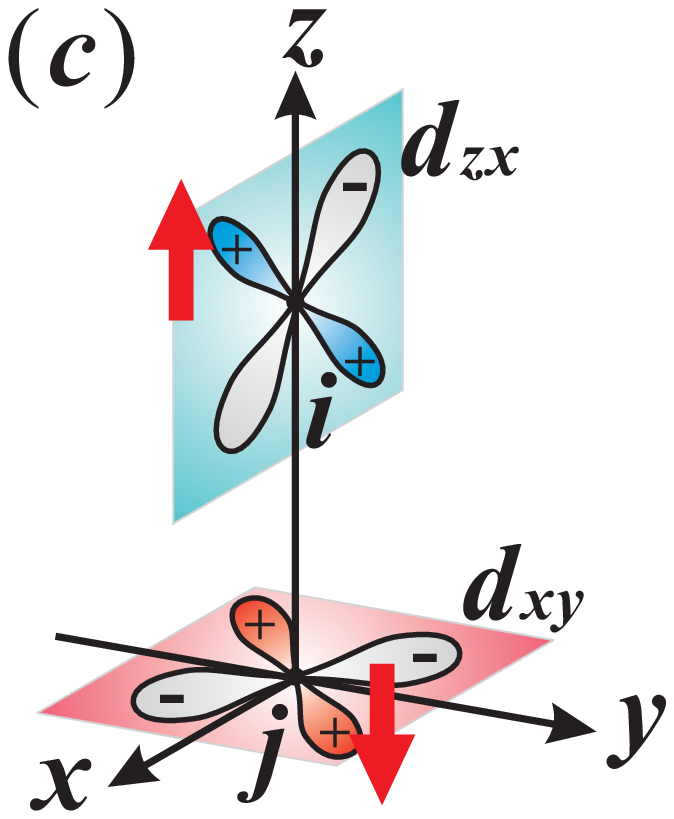,clip=1,width=0.35\linewidth,angle=0}
\centering\epsfig{file=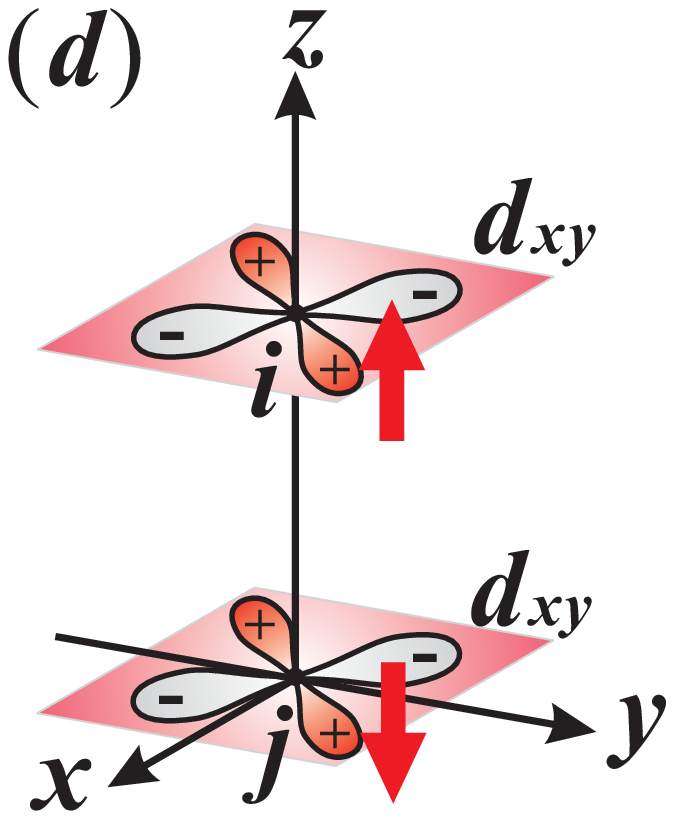,clip=1,width=0.35\linewidth,angle=0}
\caption{Representatives orbital configurations along a bond $\langle ij \rangle$ with
two orbitals at sites $i$ and $j$ initially occupied by spins $\uparrow$ and $\downarrow$ respectively.
Case (a): $[d_{zx,\uparrow}(i);d_{zx,\downarrow}(j)]$,
case (b): $[d_{zx,\uparrow}(i);d_{yz,\downarrow}(j)]$,
case (c): $[d_{zx,\uparrow}(i);d_{xy,\downarrow}(j)]$,
case (d): $[d_{xy,\uparrow}(i);d_{xy,\downarrow}(j)]$.
Any two cases of (a)-(d) are nonequivalent under the lattice symmetry transformation.
}
\label{fig:t2g_tran}
\end{figure}

Let us consider a general hole and spin configuration satisfying conditions
(I) and (II).
We pick up a bond $\avg{ij}$ and consider the $d_a$ orbital of site $i$
and the $d_b$ orbital of site $j$.
If they are occupied by spin $\sigma$ and $\sigma^\prime$, respectively,
let us denote this bond configuration as $[d_{a,\sigma}(i);d_{b,\sigma'}(j)]$.
We have the following lemma:
{\lemma
The spin configuration in $[d_{a,\sigma}(i);d_{b,\sigma'}(j)]$ can be
flipped to $[d_{a,\sigma'}(i);d_{b,\sigma}(j)]$ by
applying a series of hopping and Hund's interaction processes without finally
affecting spin and hole configurations in the rest of the system.
}

\proof
Without loss of generality, we assume the bond $\avg{ij}$ is along
the $z$ axis, and only discuss how to flip
$[d_{a,\uparrow}(i);d_{b,\downarrow}(j)]$ to $[d_{a,\downarrow}(i);d_{b,\uparrow}(j)]$.
Since $a$ and $b$ can take any of the $xy$, $yz$, and $zx$,
there are 9 possible orbital configurations for a bond.
Nevertheless, they can be classified into 4 non-equivalent classes
because of the lattice geometry as shown in Figs. \ref{fig:t2g_tran}
$(a)$ to $(d)$.

For later convenience, the single hole assisted spin-flipping in the 2D
single orbital infinite $U$ Hubbard model is reviewed
in Appendix \ref{appdx:nagaoka}, which played an important role in
the proof of 
the Nagaoka FM ground state and will be employed repeatedly below.

Class $(a)$: Let us consider $a=b=zx$. The same reasoning can also apply
to the case of $a=b=yz$.
Since two orbitals and the bond are coplanar and there is
one hole in this plane, we can directly use the result
in Appendix \ref{appdx:nagaoka} to exchange their spins
$[d_{zx,\uparrow}(i);d_{zx,\downarrow}(j)]\rightarrow
[d_{zx,\downarrow}(i);d_{zx,\uparrow}(j)]$.

Class $(b)$: Let us consider $a=zx$ and $b=yz$, \textit{i.e.},
the configuration $[d_{zx,\uparrow}(i);d_{yz,\downarrow}(j)]$.
The reasoning below also applies to the case of
$a=yz$ and $b=zx$.
Let us use another orbital, $d_{yz}$ at site $i$.
First, we assume that it is occupied since we can always move an electron from
other neighboring sites and return it back afterwards.
If it is occupied by spin-$\uparrow$, 
a familiar bond configuration
$[d_{yz,\uparrow}(i);d_{yz,\downarrow}(j)]$ appears.
As already shown in class $(a)$, their spins can be exchanged to give
an intermediate configuration
$[d_{yz,\downarrow}(i);d_{yz,\uparrow}(j)]$ for class (b).
Then, on site $i$, we have both $d_{zx,\uparrow}(i)$
and $d_{yz,\downarrow}(i)$, whose spins can be further exchanged by the $H_{J_{xy}}$ term
to become $d_{zx,\downarrow}(i)$ and $d_{yz,\uparrow}(i)$.
Combining these two steps of spin exchange, the initial configuration
$[d_{zx,\uparrow}(i);d_{yz,\downarrow}(j)]$ is flipped to
$[d_{zx,\downarrow}(i);d_{yz,\uparrow}(j)]$
and the third $d_{yz}(i)$ orbital remains spin-$\uparrow$ finally.
If the $d_{yz}(i)$ orbital is occupied by spin-$\downarrow$,
we can first apply Hund's interaction to exchange spins between the
$d_{zx}(i)$ and $d_{yz}(i)$ orbitals, and then
apply the process in class $(a)$ to further exchange the spins
between two $d_{yz}$ orbitals on sites $i$ and $j$.

Class $(c)$ contains four equivalent configurations
$a=d_{zx}$, $b=d_{xy}$; $a=d_{yz}$, $b=d_{xy}$; $a=d_{xy}$, $b=d_{zx}$;
$a=d_{xy}$, $b=d_{yz}$.
Class $(d)$ only contains one configuration $a=b=d_{xy}$.
The proof for these two classes are similar to that of class $(b)$
by combining Hund's interaction and hole's hopping.
The detailed proofs are given in Appendix \ref{appdx:flip}.
\textit{Q.E.D.}

Based on Lemma 3, we can have an important property of transitivity for
the many-body matrix in any sub-Hilbert space ${\cal H}^{S_z}$.
{\lemma\textbf{(Transitivity)} Consider the Hamiltonian matrix in
the subspace ${\cal H}^{S_z}$.
For any two basis vectors, $|\{h\}, \{\sigma\}\rangle$
and $|\{g\}, \{\alpha\}\rangle$,
there always exists a series of basis vectors
$|\{h_1\}, \{\sigma_1\}\rangle$, $|\{h_2\}, \{\sigma_2\}\rangle$,
...., $|\{h_k\}, \{\sigma_k\}\rangle$ connected with nonzero matrix
elements of $H$, such that
\bea
&&\avg{\{g\}, \{\alpha\} |H| \{h^\prime_1\}, \{\sigma_1\}}
\avg{\{h_1\}, \{\sigma_1\}|H| \{h_2\}, \{\sigma_2\}}\nn \\
&\times&...
\avg{\{h_k\}, \{\sigma_k\}|H| \{h\}, \{\sigma\}}
\neq 0.
\eea
}

\proof
Firstly, we can always apply the hopping term to $| \{h\}, \{\sigma\}\rangle$
to rearrange the locations of holes of each orbital band in each layer.
Then we reach an intermediate state $| \{g^\prime\}, \{\alpha^\prime\}\rangle$
in which the locations of holes are the same as that in
$| \{g\}, \{\alpha\}\rangle$.
Since the two states $| \{g\}, \{\alpha\}\rangle$ and
$| \{g^\prime\}, \{\alpha^\prime\}\rangle$ have the same z component of the total spin $S_z$, they only differ by
their spin configurations with a permutation of spins.

Since any permutation can be generated by exchanges,
it suffices to show as below that in $| \{g^\prime\}, \{\alpha^\prime\}\rangle$
two opposite spins in any two orbitals can be exchanged by consecutively applying hoppings and Hund's interactions without finally
affecting the configuration of the rest of the system.

If the two orbitals are on the same site, it is easy to exchange their spins
by applying Hund's interaction with $H_{J_{xy}}$ once.
If they are located at different sites,
we can always find a path of successive bonds connecting these two site,
and passing through nonempty sites (here are at most a number of
$\min(L_x,L_y,L_z)$ sites with all three orbitals empty).
Then, we can have a sequence of occupied orbitals in which every two adjacent
orbitals are located on two nearest neighbor sites.
We can exchange the two spins at two ends of this path as follows:
Following Lemma 3, we can flip different spins at occupied orbitals on two neighboring sites.
Then, by successively applying this operation, we are able to exchange the spins
of two ends without affecting other parts of the system.
\textit{Q.E.D.}

Now we are ready to prove the following theorem.
{\theorem\textbf{(3D FM Ground State)}
Consider the Hamiltonian $H_{kin}+H_{int}$ satisfying conditions
$(I)$ and $(II)$.
The physical Hilbert space is ${\cal H}^{S_z}$.
For any values of $V$ and $J>0$, the ground states are
fully spin-polarized and are unique
apart from the trivial spin degeneracy.
They can be expressed as
\bea
\ket{\Psi_G^{S_z}}={\sum}^\prime
c_{\{h\},\{\sigma\}}
\ket{\{h\},\{\sigma\}}
\label{eq:groundstate}
\eea
where all the coefficients are strictly positive and ${\sum}^\prime$
means the summation over states in ${\cal H}^{S_z}$.
}

\proof
Because of Lemma 2 of non-positivity and Lemma 4 of transitivity,
the Hamiltonian matrix within ${\cal H}^{S_z}$ satisfies the
prerequisites of the Perron-Frobenius theorem theorem.
The importance of the transitivity to the non-degenerate
ground state is also explained in Sec. III of
the supplementary material of Ref. [\onlinecite{li2014}].
Then it is straightforward to
conclude that Eq. (\ref{eq:groundstate}) is true which is
non-degenerate within each ${\cal H}^{S_z}$.

To show that $\ket{\Psi_G^{S_z}}$ in Eq. (\ref{eq:groundstate})
is a fully spin-polarized state, we introduce a reference state
in ${\cal H}^{S_z}$ by summing over all its bases with equal weights,
\bea
\ket{\Psi_{ref}^{S_z}}={\sum}^\prime \ket{\{h\},\{\sigma\}}.
\eea
Since $\ket{\Psi_{ref}^{S_z}}$ is symmetric under exchanging spins
of any two orbitals, it is a fully spin-polarized state
with the total spin $S=N_{tot}/2$ and its $z$ component $S_z$.
Apparently, $\avg{\Psi_G^{S_z}|\Psi_{ref}^{S_z}}\neq 0$.
Since $\ket{\Psi_G^{S_z}}$ is the unique ground state in ${\cal H}^{S_z}$,
these two non-orthogonal states must share the same good quantum numbers of $S$ and $S_z$.
{\it Q.E.D.}

Because of the spin $SU(2)$ symmetry, all ground states $\ket{\Psi_G^{S_z}}$ in different ${\cal H}^{S_z}$ with $-N_{tot}/2\le {S_z} \le N_{tot}/2$ are degenerate, and
form a set of spin multiplets
with the maximal total spin $S=N_{tot}/2$.

{\remark
Theorem 1 is true for both the periodic and open boundary conditions.}

Based on Theorem 1, we have the following two corollaries with their
proofs presented in Appendix \ref{appdx:corollary}.
{\corollary
Under condition I and a modified condition II: There is
one and only one doubly occupied orbital for each orbital band
in each layer; we have that the Hamiltonian of Eqs. (\ref{eq:kin}) and (\ref{eq:Hint})
also possesses the fully spin-polarized FM ground state which is unique up to
the trivial spin degeneracy.
}

{\corollary
If there is one and only one particle in each orbital band in each
layer, we also have that the ground state is fully spin-polarized
and unique up to the trivial spin degeneracy for any values of
$J>0$ and $V$.
}

\section{Ferromagnetism in the 2D $t_{2g}$ orbital layer}
\label{sect:layer}

In this section, we will consider the same multiorbital Hubbard
Hamiltonian of Eqs. (\ref{eq:kin}) and
(\ref{eq:Hint}) but in a single layer along the $xy$-plane.
The $d_{xy}$ orbital band remains 2D; while,
the $d_{zx}$- and $d_{yz}$ orbitals form crossed 1D bands with
dispersion perpendicular to each other.
The FM ground state of this system will be discussed when both 1D and 2D bands present.

When only the two quasi-1D bands are considered, the FM ground state has been
proved in Ref. [\onlinecite{li2014}]
under condition (I) and the following two conditions:\\
 \ (III): \textit{Open boundary condition or periodic (anti-periodic)
boundary condition with odd (even) number of particles in each row or column;}\\
 \ (IV): \textit{Arbitrary filling with at least one hole and one particle in each
row and each column.}

To describe the part of $d_{zx}$ and $d_{yz}$ bands with general fillings,
let us first recapture the many-body bases constructed for the quasi-1D
system in Ref. [\onlinecite{li2014}] and rewrite them
in terms of $d_{zx}$ and $d_{yz}$ bands.
By Lemma 1, for any generic filling,
we can always specify a partition of particle numbers into
rows  $\mathcal{X}=\left\{ r_i =1, \cdots, L_y\right \}$
and columns $\mathcal{Y}=\left\{ c_i =1, \cdots, L_x \right \}$ as,
$\mathcal{N}_{\mathcal{X}}=\left\{ N_{r_i} \right \},
\mathcal{N}_{\mathcal{Y}}=\left\{ N_{c_i} \right \}$,
where $N_{r_i}$ and $N_{c_i}$ are the particle numbers
of $d_{zx}$- and $d_{yz}$ orbitals conserved in the
$r_i$-th row and the $c_i$-th column, respectively.
We can order electrons in each row from the left most particle
to the right most one, followed by the ordering in each column
from the top to bottom.
The corresponding many-body basis can be set up as
\bea
&&
\ket{\mathcal{R}, \mathcal{S}}
_{\mathcal{N}_{\mathcal{X}}, \mathcal{N}_{\mathcal{Y}}}
=\prod_{j=1}^{L_x} d_{yz,c_j}^{\dagger}
\prod_{j=1}^{L_y} d_{zx,r_j}^{\dagger} \ket{0} \nn \\
&=& d_{yz,c_{L_x}}^{\dagger} \cdots d_{yz,c_2}^{\dagger} d_{yz,c_1}^{\dagger}
 d_{zx,r_{L_y}}^{\dagger} \cdots d_{zx,r_2}^{\dagger}
 d_{zx,r_1}^{\dagger} \ket{0}, \nn \\
\label{eq:1d_basis}
\eea
where $j$ denotes the index of columns and rows;
$\mathcal{R}=\{ \mathbf{r}^{r_j}_i; \mathbf{r}^{c_j}_i|$ all $i$'s and $j$'s$\}$ represents
the  coordinates of occupied sites;
$\mathcal{S}=\{ \alpha^{r_j}_i; \beta^{c_j}_i|$ all $i$'s and $j$'s$\}$ represents their
the spin configurations.
The operator $d_{zx,r_j}^{\dagger}$ ($d_{yz,c_j}^{\dagger}$) creates a whole line of $N_{r_j}$($N_{c_j}$) $d_{zx}$($d_{yz}$) electrons in the row $r_j$ (column $c_j$) ordered from left to right (from top to bottom), $d_{zx,r_j}^{\dagger} = \prod_{\mathbf{r}_i \in \textrm{row} r_j} d_{zx}^{\dagger}(\mathbf{r}_1)d_{zx}^{\dagger}(\mathbf{r}_2)\cdots d_{zx}^{\dagger}(\mathbf{r}_{N_{r_j}})$, and $d_{yz,c_j}^{\dagger}$ can be similarly defined.

Now, let us consider the additional quasi-2D $d_{xy}$ band with one and only one hole.
The basis for this layer of $d_{xy}$ orbital $|h^{xy},\{\sigma\}\rangle$
is defined following  Eq. (\ref{eq:sign}) but without the layer index.
Then, the basis for the Hilbert space of this 2D system ${\cal H}_{2D}$ can be constructed by the
direct product of the basis for the 1D bands and that for the 2D band,
\bea
\ket{\mathcal{R}, \mathcal{S}}_{\mathcal{N}_{\mathcal{X}}, \mathcal{N}_{\mathcal{Y}}}
\otimes |h^{xy},\{\sigma\}\rangle.
\label{eq:basis_RS}
\eea
Again, because of the conservation of the $z$ component of the total spin, this Hilbert space
can be decomposed as ${\cal H}_{2D}=\bigoplus {\cal H}_{2D}^{S_z}$.
Following the same steps in Ref. [\onlinecite{li2014}] and in
Sec. \ref{sect:3d}, it is straightforward to show that for the basis
defined in Eq. (\ref{eq:basis_RS}), and under condition (III) for 1D bands,
the off-diagonal matrix elements of the many-body Hamiltonian
are non-positive.

Below, we further show the transitivity of the Hamiltonian
matrix in the sub-Hilbert space  ${\cal H}_{2D}^{S_z}$
under condition (IV) for $d_{zx}$- and $d_{yz}$ bands.
Since the locations of electrons can be easily adjusted by applying hopping
terms, it suffices to show the transitivity between two bases only differ by spin configurations,
$|u\rangle=\ket{\mathcal{R}, \mathcal{S}}_{\mathcal{N}_{\mathcal{X}},
\mathcal{N}_{\mathcal{Y}}}\otimes |h^{xy},\{\sigma\}\rangle$
and $|v\rangle=\ket{\mathcal{R}, \mathcal{S^\prime}}_{\mathcal{N}_{\mathcal{X}},
\mathcal{N}_{\mathcal{Y}}}\otimes |h^{xy},\{\sigma^\prime\}\rangle$.
Again, we only need to show that for the state of $|u\rangle$,
we can exchange any two different spins by applying hopping and
Hund's interaction terms.
If these two electrons are both in quasi-1D bands $d_{zx}$ and $d_{yz}$,
this situation has been proved in Ref. [\onlinecite{li2014}].
If these two electrons are both in the $d_{xy}$ band, it is reduced
to the usual case of the Nagaoka system.

%

Now let us consider the case of one electron in the quasi-1D bands,
without loss of generality, in the $d_{zx}$ orbital band with spin-$\uparrow$; and another electron in the $d_{xy}$ band with spin-$\downarrow$.
We denote their locations as $\mathbf{r}_{zx}$ and $\mathbf{r}_{xy}$, respectively.
Let us identify the site $\mathbf{r}_c$ which is in the same row of
the $d_{zx}$ electron and in the same column of the $d_{xy}$ electron,
and consider the $d_{yz}$ orbital at this site.
We assume that there is an electron of the $d_{yz}$ orbital
at $\mathbf{r}_c$.
If not, because of condition (IV), we can always move a  $d_{yz}$ electron
of that column to $\mathbf{r}_c$ by hopping.  And the configuration in this column can be restored by reversing the hopping afterward.
If the electron of the $d_{yz}$ orbital at $\mathbf{r}_c$ has spin-$\uparrow$, it can first be moved to $\mathbf{r}_{xy}$ by hoppings. Then, it can exchange the spins with the $d_{xy}$ electron at $\mathbf{r}_{xy}$ by Hund's interaction. After reversing the hopping, this $d_{yz}$ electron can be moved back to $\mathbf{r}_c$ but with spin-$\downarrow$.
Further, it can be moved to $\mathbf{r}_{zx}$ to exchange the spins with the $d_{zx}$ electron and be moved back to $\mathbf{r}_c$ again with its original spin-$\uparrow$ recovered.
The net effect is the exchange of spin configurations
between the $d_{xy}$ and $d_{zx}$ electrons without affecting
other configurations. The case of the $d_{yz}$ electron at $\mathbf{r}_c$ with spin-$\downarrow$ can be similarly proved.

So far, we have shown both the non-positivity of off-diagonal matrix elements
and the transitivity of the Hamiltonian matrix in the sub-Hilbert
space ${\cal H}_{2D}^{S_z}$.
Then, following the same reasoning in the proof of Theorem 1,
it is straightforward to have the following theorem
{\theorem\textbf{(2D FM Ground State)}
Consider the case in which there is one and only one hole in the $d_{xy}$ band.
Under conditions $(I)$, $(III)$, and $(IV)$, for any values
of $V$ and $J>0$, the ground states are fully spin-polarized
which is unique apart from the trivial spin degeneracy.
}

Next, we consider the situation in which the $d_{xy}$ band is half-filled,
\textit{i.e.}, there is no hole.
In this case, the $d_{xy}$ band by itself is not ferromagnetic.
Because of the coupling to the quasi-1D band, we have the following
theorem
{\corollary
If the $d_{xy}$ band is half-filled, under the same conditions
in Theorem 2, for any values
of $V$ and $J>0$, the ground states are fully spin-polarized
which is unique apart from the trivial spin degeneracy.
}

\proof
We first define the basis for the local moments for the
half-filled $d_{xy}$ band, which can be ordered in an arbitrary way as
\bea
|\{\sigma\}\rangle= \prod_{i}
d^\dagger_{xy,\sigma} (i)|0\rangle,
\label{eq:sign2}
\eea
where $\avg{\sigma}$ is an arbitrary spin distribution.
Then for the combined system, the basis is defined as
\bea
\ket{\mathcal{R}, \mathcal{S}}_{\mathcal{N}_{\mathcal{X}}, \mathcal{N}_{\mathcal{Y}}}
\otimes  |\{\sigma\}\rangle.
\eea
Again because of spin conservation, the Hilbert space in this case
${\cal H}^\prime_{2D}$ can be further decomposed into the direct sum
of different sectors of $S_z$'s as ${\cal H}^{\prime}_{2D}=\oplus {\cal H}^{S_z \, \prime}_{2D}$.

Similarly to Theorem 2, the off-diagonal elements of
the Hamiltonian matrix is non-positive.
We next show the transitivity of the Hamiltonian matrix in
each physical sub-Hilbert space ${\cal H}^{S_z \, \prime}_{2D}$.
Again, we only need to show that for any state in ${\cal H}^{S_z \, \prime}_{2D}$,
opposite spins of any two electrons can be exchanged by applying
hopping and Hund's interaction without affecting other parts of the system.
The proof is very similar to that of Theorem 2,
Nevertheless, a new situation needs to be addressed: both
electrons are in the $d_{xy}$ band with spin-$\uparrow$
and -$\downarrow$, respectively.
Their locations are denoted as $\mathbf{r}$ and $\mathbf{r}^\prime$,
respectively.
Then we can choose an electron in the $d_{zx}$ band, and,
without loss of generality, assume its spin-$\uparrow$.
Then according to the proof of Theorem 2,
we can first flip the pair of electrons $d_{zx,\uparrow}$ and
$d_{xy,\downarrow}(\mathbf{r}^\prime)$, then their spins become
$d_{zx,\downarrow}$ and $d_{xy,\uparrow}(\mathbf{r}^\prime)$.
Next, we consider the pair of $d_{zx,\downarrow}$ and
$d_{xy,\uparrow}(\mathbf{r})$ and exchange their spins.
The net result is the exchange of the spins of two $d_{xy}$ electrons.

With both results of non-positivity and transitivity, it is
also straightforward to arrive at Corollary 3 by similar proof of Theorem 1. \textit{Q.E.D.}

\section{Discussion on experiments}
\label{sect:exp}

Although Theorems 1 and 2 are under ideal conditions and limits,
they do have close connections to realistic systems of transition-metal oxides.
For the multiorbital Hubbard Hamiltonian of Eqs. (\ref{eq:kin}) and
(\ref{eq:Hint}), they are actually a good approximation of the
$t_{2g}$ orbital systems of transition metal oxides in 3D.
For example, the itinerant FM SrRuO$_3$ belongs to this class of materials
\cite{koster2012,cao1997,jeong2013}, which is a $t_{2g}$-active material of
4$d$ electrons in a cubic lattice.
Even though, typical interaction strength in the 4$d$ electron systems are
intermediately strong, it already exhibits the FM phase with
$T_c=165K$. Furthermore, the magnetic moment of this system is observed as 1.6$\mu_B$ per site with the electron filling in  SrRuO$_3$ as four electrons per site.
Therefore, the FM ground state stated in Theorem 1 would possibly
persist to the intermediate interaction regime and with finite
electron or hole density away from half-filling.
Nevertheless, the magnetization would be no longer fully polarized
but partially polarized to save the kinetic energy cost.

Another important system is the LaAlO$_3$/SrTiO$_3$ interface between two component insulators.
This interface is experimentally found metallic and ferromagnetic with large magnetization
\cite{lilu2011,bert2011}.
This is a $t_{2g}$ orbital active material with 3$d$ electrons in
2D layered systems, whose $d_{zx}$ and $d_{yz}$ are quasi-1D orbital
bands while its $d_{xy}$ orbital forms the quasi-2D band.
For 3$d$ electrons, the interaction strength is stronger than
that of 4$d$ materials.
The RKKY, itinerant, and double-exchange mechanisms
were proposed to explain the FM in this system \cite{michaeli2012,chen2013,banerjee2013}.
Here, we have shown that the ground state itinerant FM
is fully spin polarized and robust for general densities
in the $d_{zx}$ and $d_{yz}$ bands under strong intra orbital
interactions.

\section{Conclusions}
\label{sect:cons}
In summary, we have investigated the Nagaoka type itinerant FM in
$t_{2g}$ orbital systems in a 3D cubic lattice.
The hole motion in each orbital band is constrained in the plane
parallel to the orbital orientation. Effectively, this system
behaves as crossing planes of 2D Nagaoka FM coupled by
on-site inter orbital Hund's coupling.
Consequently, 3D itinerant FM ground states are developed, which are proved fully polarized and unique apart from the trivial spin multiplet degeneracy. Also, we have considered the 2D layer of $t_{2g}$ orbital systems:
the quasi-1D bands are itinerant with arbitrary generic
fillings and the quasi-2D band can have a single hole or
be half-filled. Its ground state is shown remaining the fully spin-polarized itinerant FM.
The theorems established in this article can be helpful for further
understanding the mechanism of FM in SrRuO$_3$ and the transition-metal oxides interface.

\acknowledgements
Y.L. is grateful to Elliott H. Lieb and Congjun Wu for their helpful discussion and encouragement. Y.L. thanks the Princeton Center
for Theoretical Science for support.

\appendix

\section{Expressions and physical meaning of $U$, $V$, $J$ and $\Delta$}
\label{appdx:parameter}

The expressions of $U$, $V$, $J$ and $\Delta$ in Eq. (\ref{eq:Hint}) are
standard two-body Coulomb integrals under the $t_{2g}$ orbital basis.
We assume the bare Coulomb interaction as $V(\mathbf{r}_1-\mathbf{r}_2)$,
and express the Wannier $t_{2g}$ orbital wavefunctions
$\phi_{a}(\mathbf{r})$ with $a=xy,yz$ and $zx$, respectively.
Then $U$, $V$, $J$ and $\Delta$ can be represented
\cite{hubbard1963,slater1953} as
\bea
U&=&\int \textrm{d} \mathbf{r}_1  \textrm{d} \mathbf{r}_2
\phi_{a}^* (\mathbf{r}_1) \phi_{a}^* (\mathbf{r}_2)
V(\mathbf{r}_1-\mathbf{r}_2) \phi_a (\mathbf{r}_2) \phi_a (\mathbf{r}_1), \nn \\
V&=& \int \textrm{d} \mathbf{r}_1 \textrm{d}\mathbf{r}_2
\phi_a^* (\mathbf{r}_1) \phi_b^* (\mathbf{r}_2)
V(\mathbf{r}_1-\mathbf{r}_2) \nn \\
&&\times
\Big\{
\phi_b (\mathbf{r}_2) \phi_a (\mathbf{r}_1) -
\phi_a (\mathbf{r}_2) \phi_b (\mathbf{r}_1) \Big\},
\nn \\
J&=&2\int \textrm{d} \mathbf{r}_1 \textrm{d}\mathbf{r}_2
\phi_a^* (\mathbf{r}_1) \phi_b^* (\mathbf{r}_2)
V(\mathbf{r}_1-\mathbf{r}_2) \phi_a (\mathbf{r}_2) \phi_b (\mathbf{r}_1), \nn \\
\Delta &=& \int \textrm{d} \mathbf{r}_1 \textrm{d}\mathbf{r}_2
\phi_a^* (\mathbf{r}_1) \phi_a^* (\mathbf{r}_2) V(\mathbf{r}_1-\mathbf{r}_2)
\phi_b (\mathbf{r}_2) \phi_b (\mathbf{r}_1). \nn \\
\label{eq:multiband}
\eea
where $a\neq b$ and no summation over repeated indices is assumed.

Let us explain the physical meanings of $U$, $V$, $J$ and $\Delta$
by considering a single-site problem filled with only two fermions.
In total there are 15 states, which can be classified as 3 sets of spin triplets and 6 spin singlets.
The three sets of spin triplet states can be expressed as
\bea
&&d^\dagger_{a,\uparrow} d^\dagger_{b,\uparrow} |0\rangle, \ \ \, \ \ \,
d^\dagger_{a,\downarrow} d^\dagger_{b,\downarrow} |0\rangle \nn \\
&&\frac{1}{\sqrt 2} \left\{d^\dagger_{a,\uparrow} d^\dagger_{b,\downarrow}
+d^\dagger_{a,\downarrow} d^\dagger_{b,\uparrow}
\right\} |0\rangle
\label{eq:inter_trp}
\eea
with $a\neq b$, and their energy is $V$.
The 6 spin singlets can be further classified as the orbital
angular momentum (OAM) singlet, doublet and triplet as follows.
The splitting between the OAM doublet and triplet states
is because of the cubic symmetry, which is a two-particle
analogy to the single-particle version of the
$t_{2g}$ and $e_g$ level splitting.
The orbital angular momentum singlet state is expressed as
\bea
\frac{1}{\sqrt 3}\left\{d^\dagger_{xy,\uparrow} d^\dagger_{xy,\downarrow}
+d^\dagger_{yz,\uparrow} d^\dagger_{yz,\downarrow}
+d^\dagger_{zx,\uparrow} d^\dagger_{zx,\downarrow}
\right\} |0\rangle,
\eea
whose energy is $U+2\Delta$.
The orbital angular momentum doublet states have the energy
$U-\Delta$, and they are expressed as
\bea
&&\frac{1}{\sqrt 6}\left\{d^\dagger_{yz,\uparrow} d^\dagger_{yz,\downarrow}
+d^\dagger_{zx,\uparrow} d^\dagger_{zx,\downarrow}
-2 d^\dagger_{xy,\uparrow} d^\dagger_{xy,\downarrow}\right\} |0\rangle,\nn \\
&& \frac{1}{\sqrt 2}\left\{
d^\dagger_{yz,\uparrow} d^\dagger_{yz,\downarrow}
-d^\dagger_{zx,\uparrow} d^\dagger_{zx,\downarrow}\right\} |0\rangle .
\eea
The orbital angular momentum triplet states have
energy $J+V$, whose wavefunctions are expressed as
\bea
&& \frac{1}{\sqrt 2}\left\{ d^\dagger_{yz,\uparrow} d^\dagger_{zx,\downarrow}
-d^\dagger_{yz,\downarrow} d^\dagger_{zx,\uparrow} \right\} |0\rangle,\nn \\
&& \frac{1}{\sqrt 2}\left\{ d^\dagger_{zx,\uparrow} d^\dagger_{xy,\downarrow}
-d^\dagger_{zx,\downarrow} d^\dagger_{xy,\uparrow} \right\} |0\rangle,\nn \\
&& \frac{1}{\sqrt 2}\left\{ d^\dagger_{xy,\uparrow} d^\dagger_{yz,\downarrow}
-d^\dagger_{xy,\downarrow} d^\dagger_{yz,\uparrow} \right\} |0\rangle.
\label{eq:inter_sng}
\eea
Clearly, the energy difference between the inter orbital
singlet and triplet states is $J$ as comes from Hund's coupling.

\section{Spin flipping in a single orbital 2D Hubbard model
in the square lattice}
\label{appdx:nagaoka}

\begin{figure}[tbp]
\centering\epsfig{file=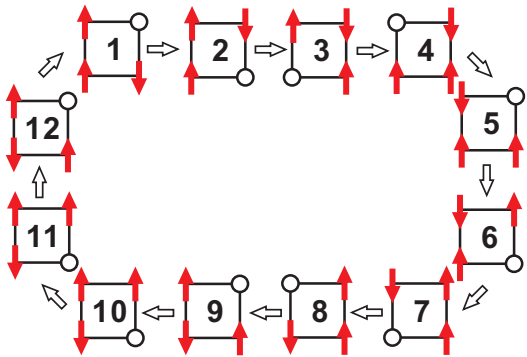,clip=1,width=0.8\linewidth,angle=0}
\caption{The 12 configurations with three electrons (2 spin-$\uparrow$
and 1 spin-$\downarrow$) and a single hole in a square plaquette.
The hole's motion builds up transitivity among all these states.
From Ref. \onlinecite{nagaoka1966}.
}
\label{fig:nagaoka}
\end{figure}

To keep this paper self-contained, we review an important step showing the transitivity in the single orbital Nagaoka system \cite{Tasaki2003}.
We only consider the case of
the 2D Hubbard model in the square lattice with $U=+\infty$
with a single hole [\onlinecite{nagaoka1966}].
The Hamiltonian can be written as
\bea
H=t\sum_{\avg{ij}} P \left\{ c^\dagger_i c_j +h.c.\right\}P,
\label{eq:kin_2}
\eea
where $P$ is the projection operator projecting out the
doubly occupied state.

Consider a bond $\avg{ij}$ with its two sites $i$ and $j$ occupied by
spins $\sigma$ and $\sigma^\prime$ in orbitals $d_a$ and $d_b$, respectively.
This configuration is denoted as $[d_{a,\sigma} (i), d_{b,\sigma'}(j)]$.
As shown below, the spins in this configuration can be exchanged to be $[d_{a,\sigma'} (i), d_{b,\sigma}(j)]$
by applying a series of hoppings in Eq. (\ref{eq:kin_2})
without affecting hole and spin configurations of other sites.

Obviously, we only need to consider the case of $s_z\neq s_z^\prime$.
Spin flipping can be realized by the following motion of the single hole.
We can choose a plaquette unit containing the bond $\avg{ij}$.
If here is a hole in this plaquette, without loss of generality,
we can assume that here are two spin-$\uparrow$'s and one spin-$\downarrow$ in
the rest 3 sites of this plaquette.
They can form in total $12$ possible combinatorial configurations.
As shown in Fig. \ref{fig:nagaoka}, they can be connected to each
other by simply applying hole hoppings clockwise in the plaquette
for at most three rounds.
Therefore, it is possible to exchange the spins on the bond $\avg{ij}$
without affecting other sites.

If this plaquette does not contain a hole, we can first apply
the hopping process to move the hole to this plaquette.
During this process, we require that the hole should not
pass sites $i$ and $j$, which is possible because even when we remove
all the bonds connecting $i$ and $j$, the remaining part of the
lattice is still connected.
Following the conclusion above, we can flip the spin configuration
on the bond $\avg{ij}$ without affecting other sites.
Afterwards, we can restore the rest of the spin configuration
by reversing the hole's motion along the same
path on which it was brought to the plaquette before.
Finally, the spin configuration on $\avg{ij}$ becomes flipped, \textit{i.e.},
$[d_{a,\sigma} (i), d_{b,\sigma'}(j)] \rightarrow [d_{a,\sigma'} (i), d_{b,\sigma}(j)]$.
Meanwhile, the hole returns to its original location
and  spin configurations on other sites are restored.

\section{Exchanging spins in Classes $(c)$ an $(d)$ }
\label{appdx:flip}
In this section, we complete the proof
of Lemma 3 for the orbital configurations of classes $(c)$
and $(d)$ below.

\proof
Class $(c)$: We consider the case of $a=zx$ and $b=xy$,
\textit{i.e.}, the configuration $[d_{zx,\uparrow}(i);d_{xy,\downarrow}(j)]$.
The reasoning below also applies to the other 3 situations
of $a=yz$, $b=xy$; $a=xy$, $b=zx$; and
$a=xy$, $b=yz$.
Here, the spin exchange between $d_{zx,\uparrow}(i)$ and $d_{xy,\downarrow}(j)$ can be aided by the $d_{zx}(j)$ orbital. Following the reasoning in the main text, $d_{zx}(j)$ can always be assumed occupied.
If it has spin-$\uparrow$, on site $j$, we have $d_{zx,\uparrow}(j)$ and $d_{xy,\downarrow}(j)$, whose spins can be exchanged by Hund's interaction to be $d_{zx,\downarrow}(j)$ and $d_{xy,\uparrow}(j)$.
Then bond $\avg{ij}$ has a new spin configuration  $[d_{zx,\uparrow}(i);d_{zx,\downarrow}(j)]$,
which can be flipped as shown in class $(a)$.
As a result, the initial configuration of $[d_{zx,\uparrow}(i);d_{xy,\downarrow}(j)]$
is flipped to $[d_{zx,\downarrow}(i);d_{xy,\uparrow}(j)]$ without
affecting $d_{zx,\uparrow}(j)$.
Similarly, if the $d_{zx}(j)$ orbital is occupied by spin-$\downarrow$,
we can first apply the process in class $(a)$ to flip the
spin configuration of $d_{zx}$ orbitals on sites $i$ and $j$,
and then apply Hund's interaction to flip spins on
$d_{zx}(j)$ and $d_{xy}(j)$ orbitals.

Class $(d)$: We consider the case in which both orbitals on $\avg{ij}$ are
transverse, \textit{i.e.}, the configuration $[d_{xy,\uparrow}(i);d_{xy,\downarrow}(j)]$.
This time we check the $d_{zx}(i)$ orbital, and first assume it is
occupied.
If its configuration is $d_{zx,\uparrow}(i)$, then along the bond $\avg{ij}$
we have $[d_{zx,\uparrow}(i);d_{xy,\downarrow}(j)]$, which can be flipped
to $[d_{zx,\downarrow}(i);d_{xy,\uparrow}(j)]$ following the steps in
class (c).
Then on site $i$, the spin configuration is changed to
$[d_{zx,\downarrow}(i);d_{xy,\uparrow}(i)]$, which can be flipped
to $[d_{zx,\uparrow}(i);d_{xy,\downarrow}(i)]$ by Hund's interaction.
As a result, the initial configuration of
$[d_{xy,\uparrow}(i);d_{xy,\downarrow}(j)]$ is flipped to
$[d_{xy,\downarrow}(i);d_{xy,\uparrow}(j)]$ and $d_{zx,\uparrow}(i)$
is maintained.
If $d_{zx}(i)$ is occupied by spin-$\downarrow$, we can first
apply Hund's interaction on site $i$ and then apply the steps presented
in class $(c)$.
Finally, if the $d_{zx}(i)$ orbital is empty, we can move this hole
to a neighboring site, and perform the above process, and
then move the hole back.

\section{The proof of Corollaries I and II }
\label{appdx:corollary}
In this part, we prove the two corollaries in Sec. \ref{sect:3d}.

(Corollary 1)
\proof We perform a particle-hole transformation, \textit{i.e.},
$d_{a,\sigma}\rightarrow d^\dagger_{a,\sigma}$.
Under this transformation, the hopping Hamiltonian Eq. (\ref{eq:kin})
remains the same except for the reversed sign of $t_\pp$.
Nevertheless, for the bipartite lattice, the sign of $t_\pp$ can be
reversed by a gauge transformation, which will not change the
physics.
The physical quantities transform as follows:
\bea
n_{a,\sigma}\rightarrow 1-n_{a,\sigma}, \ \ \,
\vec S_{a} \rightarrow -\vec S_{a}.
\eea
It is easy to check that for the interaction part $H_{int}$,
$U$, $V$, $J$, and $\Delta$ remain the same apart from
a constant and a term proportional to electron density.
In the case of fixing particle numbers, the difference
is just a constant which does not affect real physics.
Under this transformation, the doubly occupied orbitals
are mapped to holes.
According to Theorem 1, the ground states are FM states with
the total spin $S=N_{tot}/2-L_x-L_y-L_z$ and are unique up
to spin degeneracy.

(Corollary 2)
\proof We order the $d_{xy}$ electrons layer by layer and define
\bea
|\{r_{xy}\},\{\sigma\}_{xy}\rangle =\prod_{l_z=1}^{L_z}
d^\dagger_{xy,\sigma}(\mathbf{r}^{xy}_{l_z},l_z)\ket{0},
\eea
where $\mathbf{r}_{xy}$ is the in plane location of the electron
in the $l_z$-th layer.
Similar bases can also be defined for $d_{yz}$ and $d_{zx}$ electrons
as $|\{r_{yz}\},\{\sigma\}_{yz}\rangle$
and $|\{r_{zx}\},\{\sigma\}_{zx}\rangle$, respectively.
The many-body bases for the entire system can be defined as
\bea
|\{r\}, \{\sigma\}\rangle &=& |\{e_{xy}\},\{\sigma\}_{xy}\rangle
\otimes |\{e_{yz}\},\{\sigma\}_{yz}\rangle \nn \\
&&\otimes |\{e_{zx}\},\{\sigma\}_{zx}\rangle,
\label{eq:basis_2}
\eea
where $\{r\}$ and $\{\sigma\}$ represent the distributions
of electron coordinates and spins in each orbital band
in each layer.

We also need to perform a gauge transformation to flip the sign of
$t_\pp$ to be negative.
Then in this case, the off-diagonal matrix elements of hopping
are negative because hopping does not change the ordering
of electrons in the definition of Eq. (\ref{eq:basis_2}).
Because each orbital bands of each layer only contains one electron,
only $J$ and $V$ terms contribute.
Again the off-diagonal matrix elements arise from $J$, which are
also negative.

Next, we show the transitivity.
Since we can also move the positions of electrons freely, we only need to
consider two bases with the same electron locations but different
spin configurations, denoted as $|\{r\}, \{\sigma\}\rangle$ and
$|\{r\}, \{\sigma^\prime\}\rangle$.
Then, it suffices to show that for any two electrons in
$|\{r\}, \{\sigma\}\rangle$, we can flip their spin configuration.
If these two electrons live in different orbitals, say, $d_{xy}$
and $d_{yz}$, then the planes of their motions cross and share a common line
parallel to the $y$ axis.
We can move these two electrons to any site of this line, and
then apply Hund's interaction to flip their spins, and then move
back to their original locations.
If these two electrons live in the same orbital with opposite spins,
say, two $d_{xy}$ electrons but in two parallel layers.
Then we can find another electron in $d_{zx}$ orbitals.
We first choose the $d_{xy}$ electron with the spin opposite
to that of $d_{zx}$, and switch their spins.
Then combine the new configuration of the $d_{zx}$ and the
other $d_{xy}$ electron, and switch their spins.
The net effect is that two $d_{xy}$ electron spins
are flipped, and the $d_{zx}$ electron spin is restored.

Having proved non-positivity and transitivity, we can follow
the same steps in Theorem 1 to prove this corollary,
which will not be repeated here.

%

\end{document}